# Uncommon 2D Diamond-like Carbon Nanodots Derived from Nanotubes: Atomic Structure, Electronic States and Photonic Properties

D.W. Boukhvalov[1,2*], D.A. Zatsepin[2,3], Yu.A. Kuznetsova[2], V.I. Pryakhina[2], A.F. Zatsepin[2]

[1] College of Science, Institute of Materials Physics and Chemistry, Nanjing Forestry University, Nanjing 210037, P. R. China.

[2] Ural Federal University, Mira Street 19, Ekaterinburg 620002, Russia

[3] Institute of Metal Physics, Russian Academy of Sciences, Ural Branch, Yekaterinburg, Russia 620108.

**Abstract:** In this article, we report the results of relatively facile fabrication of carbon nanodots from single-walled and multi-walled carbon nanotubes (SWCNT and MWCNT). The results of X-ray photoelectron spectroscopy (XPS) and Raman measurements show that the obtained carbon nanodots are quasi-two-dimensional objects with a diamond-like structure. Based on the characterization results, a theoretical model of synthesized carbon nanodots was developed. The measured absorption spectra demonstrate the similarity of the local atomic structure of carbon nanodots synthesized from single-walled and multi-walled carbon nanotubes. However, the photoluminescence (PL) spectra of nanodots synthesized from both sources turned out to be completely different. Carbon dots fabricated from MWCNTs exhibit PL spectra similar to nanoscale carbon systems with $sp^3$ hybridization and a valuable edge contribution. At the same time nanodots synthesized from SWCNTs exhibit PL spectra which are typical for quantum dots with an estimated size of ~ 0.6–1.3 nm.

E-mail: danil@njfu.edu.cn

## 1. Introduction

A huge number of experimental and theoretical articles as well as dozens of reviews are devoted to carbon nanodots (CNTs) (see, for example, references to experimental [1–3] and theoretical studies [4]). Such interest is due to the great prospects and importance of practical application of CNDs in various promising areas, such as photonics, optoelectronics, sensorics, quantum informatics, etc. At the same time, a significant role in the formation of the structural and morphological characteristics of CNDs and, as a consequence, their functional properties are played by the method of nanodots synthesis (so called *top-down* or *bottom-up*), as well as the type of raw feedstock [5].

Some articles report the synthesis of CNDs based on polymers [6], while others report the presence of disordered carbon nanoparticles [7-10]. However, most of the articles convincingly prove the presence of some layered structures (see, e.g. Refs. [11-13]). High Resolution Transmission Electron Microscopy (HRTEM) data usually indicates the presence of some nanoparticles with different linear patterns. A number of articles report that these patterns coincide with those observed in graphite with particle sizes of 0.24 and 0.33 nm [14, 15]. The same articles report the formation of nanodiscs with a height of ~1.2–1.5 nm and a diameter of more than 3 nm. At the same time, HRTEM images of these nanodiscs usually show linear patterns no larger than 0.21 nm, which is somewhat unusual for graphene and graphite [16,17].

On the one hand, the Raman data on systems with described nanodiscs are similar to the Raman data on few-layer graphene [18]. At the same time, X-ray diffraction measurements of such objects usually show a diffuse peak, which corresponds to the presence of some graphite-like structures among the CND synthesis products [10]. On the other hand, X-ray Photoelectron Spectroscopy (XPS) [19] indicate the ratio of the contributions of carbon atoms with $sp^2$- and $sp^3$- hybridization from 10:1 [20] to 1.5:1 [21] or even 1:7 [22]. It should be noted that similar synthesis methods can lead to the fabrication of CNDs with different Raman signatures and the ratio of carbon atoms with $sp^2$- and $sp^3$-hybridization. For example, the preparation of CNDs from graphene oxide in one case can lead to the formation of samples which have carbon atoms with a $sp^2$: $sp^3$ ratio of 2:1 [23], and in another case, with a ratio of 20:1 [24]. Note that all CNDs described contain either zero nitrogen concentration or trace amounts.

Thus, despite the abundance of published experimental reports with detailed descriptions of synthesis methods and product characteristics, neither atomic structure of CNDs, nor the relationship between manufacturing methods and experimentally observed properties, such as Raman spectra and XPS (for details, see Refs. [4, 25]) have not been previously established. The sensitivity of the synthesis to a variety of manufacturing conditions limits the further industrial manufacturing of CNDs with preassigned atomic structure and physicochemical properties. In order to identify unambiguous relationships between carbon source, synthesis methods and yielding products, one should choose some simple system and synthesis method for testing. Another problem with CND is that most of the reported CNDs do not exhibit optical properties typical for quantum dots (QDs). Moreover, most of the reported CNDs demonstrate optical properties which are typical for extended (continuous) objects with defects. The described situation is typical for both *top-down* [5-7,10] and *bottom-up* [1,15,17,22] approaches to the fabrication of CNDs. Hence it follows that the search for methods for the controlled synthesis of CNDs with QD-properties is also challenging task.

Despite hundreds of papers (see review in Ref. [4]), theoretical modeling alone cannot fill the gap between knowledge about manufacturing methods and the atomic structure of final products. The standard model of CNDs is nanographene (single-layer [25,26], sometimes few-layer graphene [27] with a diameter of ~1–2 nm), sometimes with active centers at the edges [28, 29]. Unfortunately, these theoretical models do not agree with experimental data and demonstrate a lack of predictive power (for details, see our recent study reported in Ref. [25]). Examples of such a discrepancy between theory and experiment are complex fabrication methods (especially *bottom-up*) and ignoring the results of Raman and especially XPS data in theoretical calculations. Therefore, to understand the nature of the synthesis-structure-property triad, some unambiguous source of carbon and a simple method for obtaining CNDs are needed.

In the present study, well-characterized single-walled and multi-walled carbon nanotubes (S- and MCNTs) in aqueous solution are used as a carbon source. The chosen manufacturing method is also simple and unambiguous – it is laser ablation. Carbon sources and yielding products of laser treatment were characterized by microscopy, Raman and X-ray photoelectron spectroscopy. Theoretical models of atomic and electronic structure of the fabricated CNDs were merged with results of experimental measurements and employed to interpret the spectra of optical absorption, excitation, and photoemission.

## 2. Experimental and theoretical methods

Laser fragmentation of carbon nanotubes in deionized water was performed using a pulsed ytterbium fiber laser (wavelength 1064 nm, pulse duration 100 ns, maximum pulse energy 1 mJ). After laser fragmentation, a drop of aqueous solution was dried on a silicon substrate. The visualization of the initial nanotubes and particles deposited on the substrates was carried out using a Carl Zeiss Evo LS10 scanning electron microscope. Raman spectra were recorded employing a WiTec Alpha 300AR confocal Raman microscope in backscattering geometry with 488 nm excitation laser and 100x objective. Spectral resolution was not worth than 1.2–3.2 cm$^{-1}$.

The optical absorption spectra of the used solutions were recorded employing an Agilent Cary 5000 double-beam spectrophotometer in the range of 200–800 nm with a step of 1 nm using quartz cuvettes.

Carbon 1s core-level analysis was made with the help of a ThermoScientific *K*-alpha Plus XPS spectrometer. This spectrometer has monochromatic microfocused Al *K*α X-ray source and has 0.05 at. % element sensitivity. An operating pressure in analytic chamber during measurements was not worse than $1.3 \times 10^{-6}$ Pa. Dual-channel automatic charge compensator (GB Patent 2411763) was applied to exclude the charging of our sample under XPS analysis because of the loss of photoelectrons. Pre-run up procedures performed included standard degassing of the sample and analyzer binding energy scale inspection and re-calibration (if needed) employing sputter cleaned Au ($4f_{7/2}$ band), Ag ($3d_{5/2}$ band), and Cu ($2p_{3/2}$ band) inbuilt XPS Reference Standards according to ISO 16.243 XPS International Standard and XPS ASTM E2108-00 Standard.

Optical absorption spectra of the final samples were obtained using a PerkinElmer Lambda 35 spectrophotometer. The emission and excitation spectra were recorded using an experimental complex based on the Horiba Fluorolog 3 spectrometer. The input monochromator provided an excitation wavelength setting accuracy of less than 0.1 nm. In order to record emission spectra, a Synapse CCD camera was used, which provides a signal-to-noise ratio not worth than 20000/1.

Theoretical modeling was carried out using SIESTA pseudopotential code [30] employing the generalized gradient approximation (GGA-PBE) [31] for the exchange-correlation potential in a spin-polarized mode. The van der Waals correction [32] was also taken into account for the correct description of noncovalent interactions. A full optimization of the atomic positions was carried out during which the electronic ground state was consistently found using norm-conserving pseudopotentials [33] with cut-off radii of oxygen, carbon, and hydrogen as 1.15, 1.20, and 1.25 au, respectively. This was made for the cores with a double-$\xi$-plus polarization basis for the localized orbitals of non-hydrogen atoms and a double-$\xi$ for hydrogen atoms. The forces and total energies were optimized with an accuracy of 0.04 eV Å$^{-1}$ and 1.0 meV/cell (or less than 0.02 meV/atom), respectively.

The standard DFT simulation was implemented by applying a computational code that underestimates the bandgap values [34]. This disadvantage of the DFT-based methods can be fixed by using the DFT+U approach for the systems with transition metals [35], GW approximation approach (Green's function, screened Coulomb interaction) [34], using so-called hybrid functionals [36], or by estimation of the real bandgap from that calculated within the standard DFT framework [25,37]. Since, at the current level of hardware development, the calculation of hundreds of atoms by GW-methods is prohibitively computationally costly, and hybrid functionals are not implemented in the SIESTA code, we applied the third approach. In the mentioned above article [34], van Schilfgaarde *et al*. presented the results of calculations within the framework of

DFT-GGA and GW, as well as the corresponding experimental values of the bandgap of dozens of compounds in a relatively wide energy range from zero to 7 eV.

Results of the calculations performed demonstrate linear relationships between the values of the bandgaps calculated within DFT-GGA framework and measured values. Based on the data from Ref. [34], the simple semi-empirical formula has been derived in our recent work [37]:

$$E_{expt} \approx 0.8\,eV + 1.1 E_{GGA} \qquad (1)$$

We will use this formula to estimate the values of the bandgaps based on our DFT-GGA results. Note that unlike materials with precisely known atomic structure, such as single crystals or fullerenes, each carbon nanodot has some differences in atomic and, hence, electronic structure (even when the final products are fairly homogeneous). This difference corresponds to the broadening of absorption and emission peaks (see figures in Subsection 3.3). Thus, numerical estimation of the energy levels in CQDs, performed according to Equation (1), is within the range of inexactness of experimental measurements. The difference between calculated and estimated values of the energy gap also used for "scissors operator" in calculation of absorbance spectra.

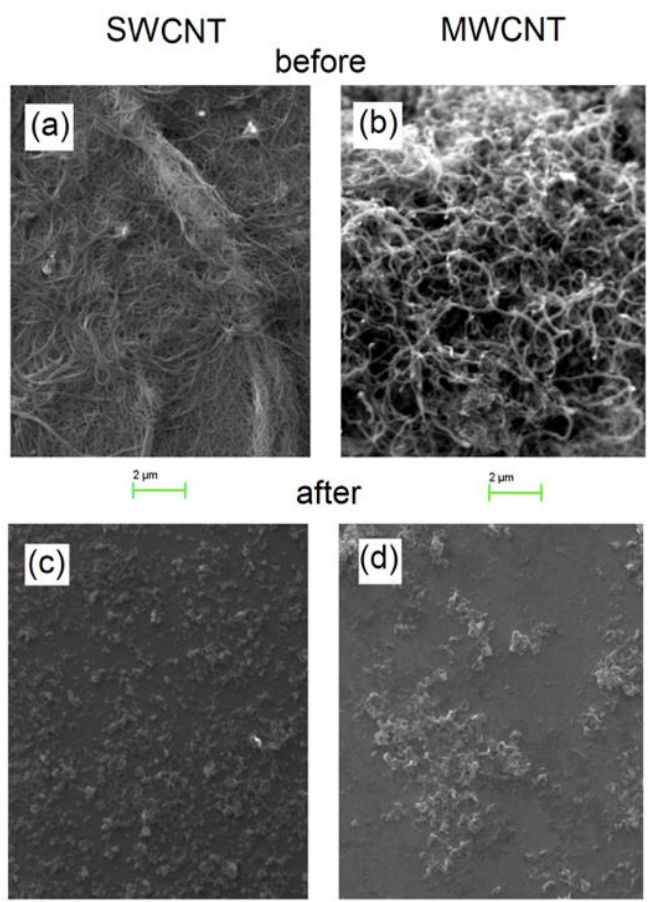

**Figure 1.** SEM images of SWCNT (a,c) and MWCNT (c,d) before (a,b) and after (c,d) laser treatment.

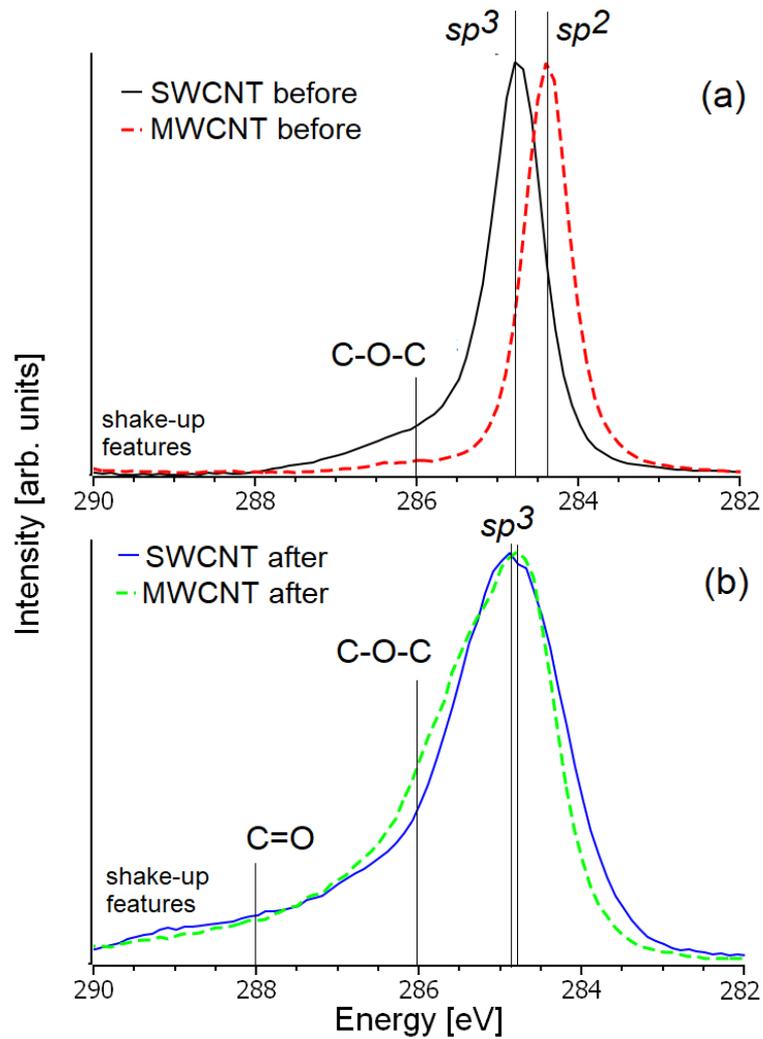

**Figure 2.** X-ray photoelectron C *1s* core-level spectra of MWCNT and SWCNT before (a) and after (b) laser treatment.

## 3. Results and Discussions

### 3.1. Structural Characterization

Scanning Electron Microscopy (SEM) measurements were carried out in order to evaluate changes in the morphology of carbon nanotubes. The results of SEM measurements are shown in Fig. 1. Based on presented data, it can be concluded that laser processing provides cutting of both single-walled and multi-walled carbon nanotubes into smaller pieces with sizes less than 100 nm. Hence, we can conclude that laser ablation provides not only a change in the local atomic structure of CNTs in the form of defects, but also the cutting of nanotubes with the formation of multiple nanosized objects.

In order to evaluate transformations of CNTs atomic structure after laser treatment, XPS measurements of the systems under study were carried out "before" and "after" laser ablation. The XPS data obtained show that SWCNTs collapsed even before treatment, and C–C bonds were also formed between the carbon planes. On the contrary, the predominance of the contribution of carbon atoms with $sp^2$ hybridization indicates the stability of MWCNTs with a local graphite-like structure. After laser treatment of the samples under study, XPS C $1s$ spectra of both S- and MWCNTs became similar. The cutting of CNTs into nanoscale fragments in an aqueous medium provides an increase in the contribution of oxidized edges (see C=O features associated with carboxyl and carbonyl groups on the edges) [10]. Another significant change in C $1s$ spectra is an increase in the contribution of oxidized carbon atoms in the basal planes (see C–O features) and a change in the main type of hybridization of MWCNTs carbon atoms from $sp^2$ to $sp^3$. Based on these results, one can assume that the energy supply to the CNTs surface in an aqueous medium is responsible for surface oxidation, which leads to the formation of 2D diamond-like structures (so-called *diamene*), previously synthesized by various methods [38–44].

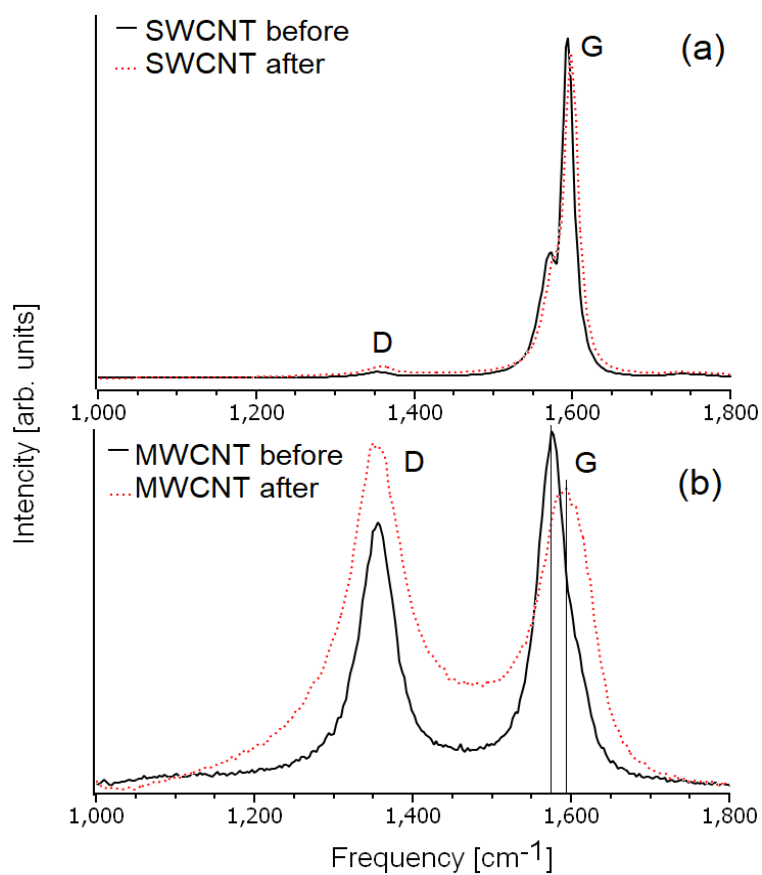

**Figure 3.** Raman spectra of SWCNT (a) and MWCNT (b) before and after laser treatment.

Raman measurements were performed in order to inspect the proposed interpretation of XPS data obtained. The results of Raman measurements (see Fig. 3) show that before treatment, both S- and MWCNTs had Raman spectra with D- and G-peaks which are typical for carbon materials with a planar (layered) structure [45]. We have to note, that these specific peaks, corresponding to ordered planar structures in carbon systems, are also present in Raman spectra of materials after treatment. Thus, we can rule out the interpretation of the $sp^2$ to $sp^3$ transition in treated MWCNTs as a result of amorphization, diamondization, or nanotube decay. The position of G-peak can provide information about the number of layers and the nature of interlayer bonds. Laser ablation leads to a negligible shift of G-peak in SWCNT-based samples (from 1609 to 1612 cm$^{-1}$), in contrast to a significant shift of this peak (more than 30 cm$^{-1}$) in MWCNT-based samples. Raman measurements of diamene show an increase in the frequency corresponding to G-peak as a result of interlayer covalent C–C bonds formation. In this case, it turns out that a larger number of layers involved in the formation of these interlayer bonds corresponds to a larger shift of G-peak [38–44]. Therefore, laser treatment of already collapsed SWCNs with a predominance of covalent interlayer bonds (see Fig. 2a) provides only a slight shift of G-peak, which may be due to the additional formation of a minority of interlayer covalent C–C bonds. This interpretation can also be supported by the decay of a low-intensity peak at 1583 cm$^{-1}$ in the SWCNT spectra after laser ablation. This peak almost coincides with G-peak in the original pristine graphene (1585 cm$^{-1}$) [46], and the decrease in this peak can be associated with a decrease in the number and size of graphene-like areas on the surface of collapsed SWCNTs. On the contrary, laser treatment of MWCNTs provides the transformation of interlayer C–C bonds from non-covalent to covalent (see Fig. 2b), which leads to a shift of G-peak, similar to that observed as a result of the transformation of several layers of graphene into diamene [38–44].

Another visually distinct peak in Raman spectra of studied samples is D-peak. This peak is usually associated with the defects such as perforations in graphene plane. Treatment of graphene with an ion beam leads to an increase in the intensity of this peak [47]. Similarly, laser ablation of carbon nanotubes leads to an increase in this peak in both SWCNTs and MWCNTs. Note that the number of vacancies and perforations in the studied SWCNTs is insignificant in contrast to the majority of these defects in MWCNTs (see Fig. 3). Here, one can assume that the formation of some perforations in SWCNTs leads to the rupture of nanotubes. On the contrary, the multiple layers in the MWCNTs prevent the collapse of the tubes after formation of some perforations in several layers. The presence of intense D-peak in the treated MWCNTs suggests that the yielding nano-objects are large enough (see also Fig. 1d) to be a stable host of a significant number of vacancies and perforations without decay or amorphization. Note that beyond highly ordered systems such as graphene and graphite [45], Raman spectra provides only qualitative information about layered carbon structures. Raman spectra measured for activated carbon, reduced graphene oxide [48, 49], fullerenes and carbon nanodots [50], carbon deposit [51] and carbon nanocones [49] sow the same broad peaks with different intensities. Detailed discussion of the limitations of Raman spectroscopy in the area of layered carbon systems can be found in the review by A. Merlen et. al., [49].

Based on the characterization results, it can be concluded that carbon nano-objects fabricated by laser ablation of CNT are ordered layered structures with predominant $sp^3$-hybridization of carbon atoms, and then a smaller number of carbon atoms have $sp^2$-hybridization, oxidized edges of nano-objects, and perforations.

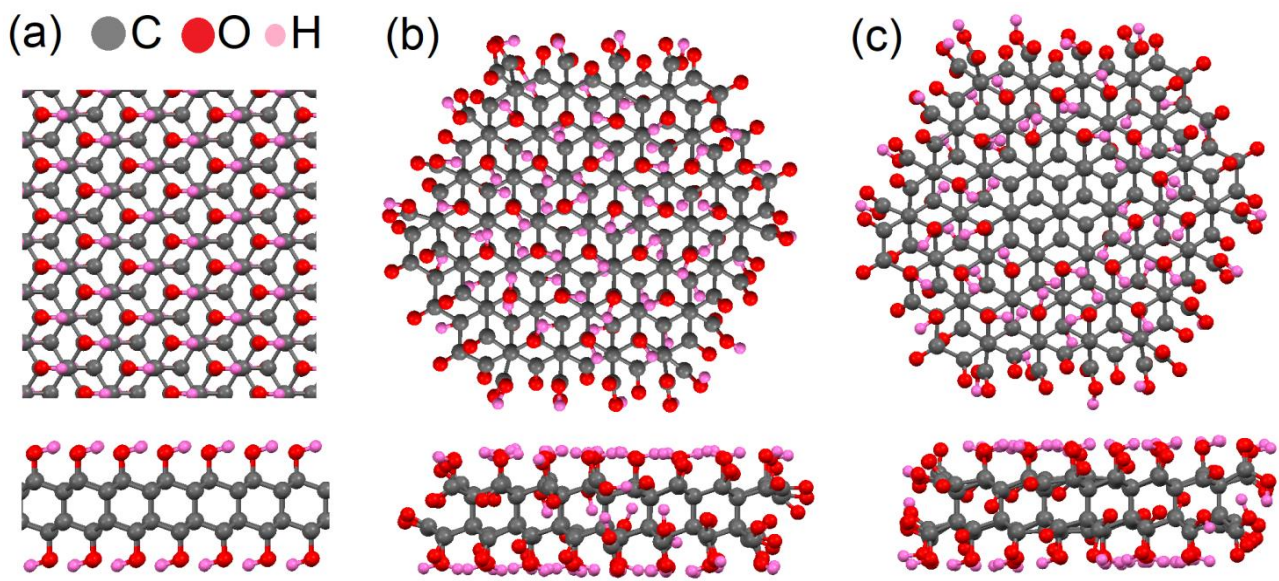

**Figure 4.** Top and side views of the optimized atomic structure of $C_2OH$ (a), infinite slab (b) unreduced (c) and maximally reduced disc (dot).

## 3.2. Theoretical Model

The first step in our modeling was to propose a model of layered carbon systems with covalent C-C bonds between the layers. A two-dimensional diamond-like slab (diamene) satisfies well these criteria, derived from abovementioned interpretation of XPS spectra and Raman scattering spectra. The transformation of bilayered graphene into a diamene can occur due to the formation of interlayer covalent C–C bonds between half of the atoms in each layer. This process corresponds to the breaking of C=C double bonds in the layer and appearance of dangling bonds with an unpaired electron on the other part of the layer (for details, see [52]). Note that the functionalization of half of the carbon atoms in each layer of the set of graphene bilayers leads to the formation of interlayer covalent carbon bonds [53]. More recent approach has been used in order to prepare diamene by hydrogenation or fluorination of the surface of several graphene layers [38-45]. Note that the removal of adatoms from the surface ensures the breaking of interlayer covalent C–C bonds and a return to initial graphene structures [39, 45]. It is noteworthy that the described processes can also be implemented for several layers of graphene [38, 39]. In the case of SWCNTs collapse, similar processes occur with the formation of covalent bonds between the layers and appearance of dangling bonds on the surface. It turns out that the saturation of these dangling bonds is extremely energetically favorable [52]. Since SWCNTs collapse occurs in liquid media and the attachment of hydroxyl groups to graphene plane is very energetically favorable [54], we will use hydroxyl groups to functionalize the 2D diamond surface (see Fig. 4a). In the case of MWCNTs, two scenarios can be implemented with yielding similar products. The first scenario: laser ablation could provoke the collapse of nanotubes, followed by addition of hydroxyl groups to the surface, similarly to that described above for SWCNTs. Conversely, in the second scenario, laser treatment provides the decomposition of water with functionalization of surface by hydroxyl groups, which ensures the formation of interlayer covalent C–C bonds. Note that the second scenario is similar to the processes used to manufacture 2D diamonds [38–44].

Based on the results of measurements and discussion above, we propose an infinite diamond-like slab passivated with hydroxyl groups (Fig. 4a) as the initial model of obtained carbon nanosystems (see details in Ref. [25]). All carbon atoms in this slab have $sp^3$-hybridization and the chemical formula of such a compound is $C_2OH$. Half of carbon atoms have four C–C single bonds, and the other half of carbon atoms have three C–C bonds and one C–O bond. Hence the ratio of C–C and C–O bonds in this system is about 7:1. This ratio is in qualitative agreement with XPS C *1s* core-level spectra (see Fig. 2b). The electronic structure of $C_2OH$ (as well as the atomic one) is also close to the electronic structure of diamond (see Fig. 5a). The value of energy gap estimated according to the Equation (1) for the structure shown in Fig. 4a is about 4.1 eV (302 nm).

The next step in our simulation was to take into account the finite size of the 2D system under study and the contribution of large perforations in the samples manufactured from MWCNTs. Based on XPS data obtained (Fig. 2b), previous studies of carbon nanotube unzipping [55,56], graphene edge passivation [57] and hence the edges of larger perforations, we propose a model structure of 368 atoms shown in Fig. 4b. The central part of this structure is identical to $C_2OH$ slab shown in Fig. 4a, and double and single dangling bonds at the edges are passivated with carbonyl and carboxyl groups, respectively. The ratio of C–C:C–O:C=O bonds in this system is 14:2:1, which is in good agreement with XPS data obtained (Fig. 2b). The transition from infinite to finite $C_2OH$ system provides a decrease in the band gap (see Fig. 5a). The value of band (or HOMO-LUMO) gap estimated using Equation (1) is about 1.45 eV (855 nm).

The last step in our modeling was to include some additional $sp^2$-hybridized carbon atoms in our model. In order to simulate these atoms, we removed one hydroxyl group from one ("upper") side of the $C_2OH$-based disc and one hydroxyl group located from the other ("lower") side of the same system. Then the number of removed hydroxyl groups was gradually increased to twelve (six on each side). It turned out that the formation of a reduced area in the central part of simulated nanodisk does not change the overall atomic structure of system (see Fig. 4c).

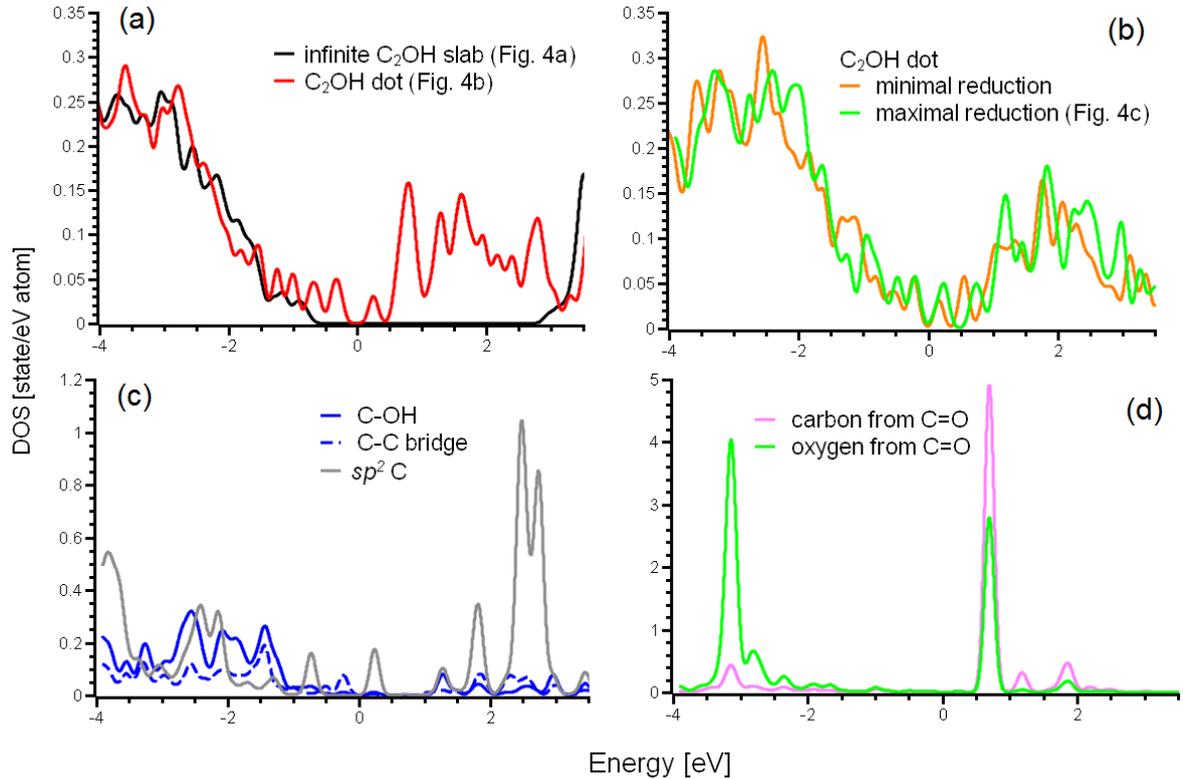

**Figure 5.** Total Densities of States of $C_2OH$ for (a) infinite diamene slab (Fig. 4a) and unreduced dot (Fig. 4b); (b) is the same dot with minimal and maximal considered level of reduction shown on Fig. 4c. Densities of States of different carbon atoms in central part (c) and carbon and oxygen atoms from C=O groups (d) on the edges of model system in Fig. 4c. Fermi level was set as zero.

In order to inspect the effect of partial reduction of $C_2OH$ nanodisks on the final electronic structure, we calculated Densities of States for the minimum and maximum number of removed pairs of hydroxyl groups. Calculation results (see Fig. 5b) show that partial reduction of $C_2OH$-based dots does not lead to significant changes in the electronic structure and band gap of the system under study. In order to estimate the contribution of different atoms to the electronic structure, we visualized the Densities of States of these atoms in the structure shown in Fig. 4c. The results of our calculations demonstrate that carbon atoms with $sp^3$-hybridization are practically excluded from the formation of states near the Fermi level (see Fig. 5c). Opposite $\pi$-orbitals of carbon atoms from reduced area in the central parts of the model dot shown in Fig. 4c form distinct peaks near zero (see Fig. 4c). Edge groups, such as C=O, also make a visible contribution to the area near the Fermi level (Fig. 4d), as discussed earlier for nano-graphenes [58]. Consequently, optical properties of the systems under study do not depend on the exact size and shape of reduced areas or dots, but are determined by the contributions of $\pi$-orbitals of carbon atoms with *sp²*-hybridization from the reduced areas or C=O groups at the edges. To illustrate this conclusion, optical absorption spectra of the structures shown on Fig. 4 have been calculated. Results of the calculations shown in Fig. 6a demonstrate that turn from $C_2OH$ infinite sheet to the nanodisc leads to dramatic changes in absorption spectra. We see that the calculated optical absorption spectra (Fig. 6a) for used structural models of nanodots is in agreement with experimental data (Fig. 6b). It should also be noted, that inhomogeneity of the structure shown on Fig. 4c related with introducing of *sp²*-hybridized area can provide sharpening of the peaks locates between 300 and 400 nm in calculated optical absorption spectra. A detailed discussion of the nature of the observed absorption peaks and the features of the luminescent properties of nanodots is given in the next Section 3.3.

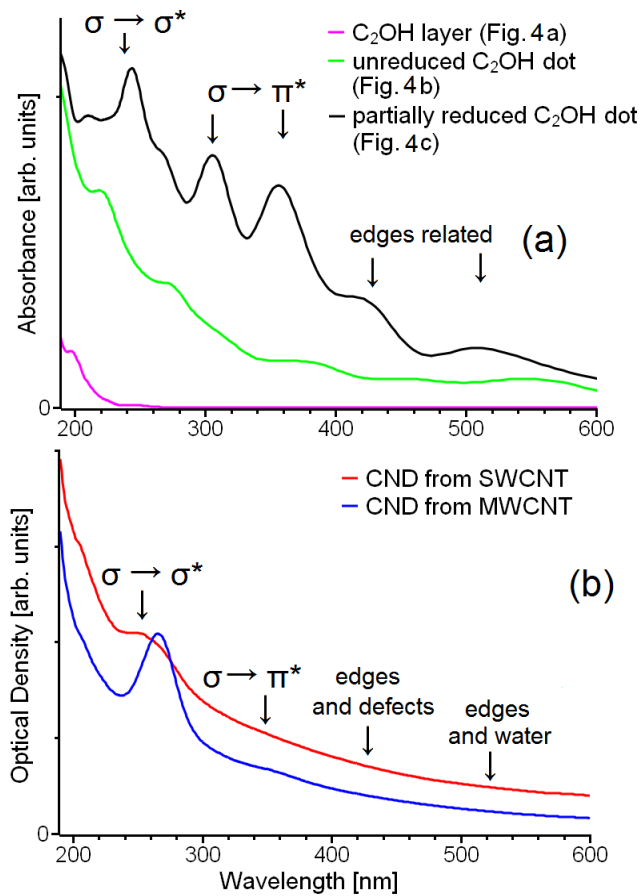

**Figure 6**. Comparison of calculated and experimental optical absorption spectra of CDs. (a) - Calculated absorbance for model structures; (b) - experimental optical absorption spectra of CDs prepared from MWCNT and SWCNT. Vertical arrows indicate the bands corresponding to $\sigma \to \sigma^*$ and $\sigma \to \pi^*$ transitions in diamond-like nanodots and transitions in defects related with edges carbonyl and carboxyl groups.

### 3.3. Optical absorption and photoluminescence

The next stage of our research is to determine the effect of the precursor (MWNT and SWCNT) on the optical properties of the obtained carbon nanosystems. First, the optical absorption spectra of S- and MWCNTs were measured after sample's treatment. The measurement results (see Fig. 6b) indicate the similarity in the absorption of both systems and rather good agreement with calculated spectra of $C_2OH$ nanodots. The dominant absorption peak at about 250–265 nm and a weak absorption peak at 350 nm can be associated with the σ → σ* and σ → π* transitions in the partially reduced obtained diamene-like nanostructures. The tail of absorption spectra, corresponding to wavelengths above 400 nm, is associated with near-gap local defect and edges related states (optical transitions *def* → *def*\*, as we noted), as well as with absorption by water associated with hydrophilic groups on the surface and edges [59]. The mentioned local defect states usually arise from functional groups of various types localized on the surface of nanodots [60]. Since experimental spectra represent superposition of the spectra of CND with some differences in atomic structures, the features in absorption spectra related with defects and edges is less distinct than in the theory (Fig. 6a). As the XPS data show (see Fig. 2b), in our case there are oxygen-containing functional groups (namely carbonyl and carboxyl groups). Note that the ratio of absorption peaks corresponding to the σ → σ*, σ → π*, and *def* → *def*\* transitions is similar to that for the contribution of C–C bonds between carbon atoms in $sp^3$-hybridization (only sigma orbitals) and another contribution to XPS core-level spectra (Fig. 2b). The similarity of absorption spectra of carbon nanodots obtained from S- and MWCNTs indicates the similarity of atomic structure of both groups of compounds under study.

Next, we considered the photoluminescent properties of CDs fabricated from S- and MWCNTs. Since PL properties of CNDs fabricated from S- and MWCNTs are very different, we will discuss these two classes separately. Figure 7a shows photoluminescence spectra of CDs prepared from MWCNTs. These spectra were recorded at different excitation wavelengths. The luminescence spectrum contains two bands with maxima at 525 nm and 558 nm. A distinctive feature of emission spectra is their constant spectral position with a variation of excitation wavelength. Thus, in CNDs manufactured from MWCNTs, the so-called excitation-independent luminescence is realized [61]. The nature of excitation-independent luminescence in CDs is usually associated with radiative transitions in surface functional groups of various types. At the same time, the spectral composition of emission can vary from the blue to the red range and is determined by the type of solution in which the CNDs were stabilized [62]. In our case, for CDs stabilized in an aqueous solution, a green emission is observed, which is typical for oxygen-containing functional groups formed on the CNDs surface, as was shown in a number of articles (see, i.e., Refs. [63, 64]). The XPS data obtained (Fig. 2) confirm oxidation of carbon atoms after laser treatment with the formation of carbonyl and carboxyl groups (Fig. 4b). Note that for CNDs fabricated from MWCNTs, the luminescence spectra do not show size effect, which consists in excitation-dependent emission [65]. Thus, it can be assumed that laser treatment of MWCNTs leads to the formation of rather large CNDs with dimensions of more than tens of nanometers. This is in a good agreement with Raman data indicating the formation of quite large nano-objects during the treatment of MWCNTs (see Section 3.1).

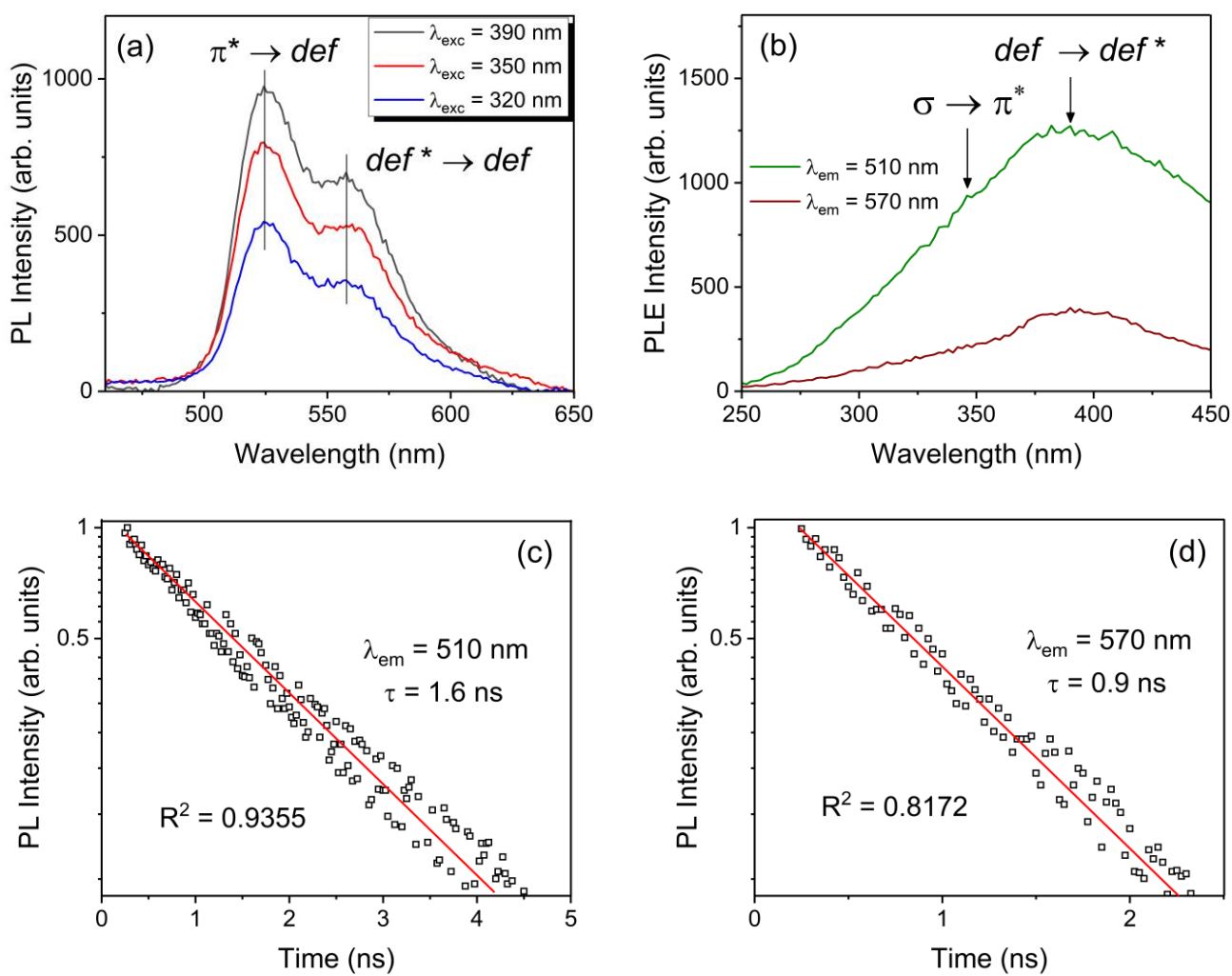

**Figure 7.** Photoluminescence (a) and excitation (b) spectra of CQDs fabricated from MWCNTs; decay of emission kinetics at 510 nm (c) and 570 nm (d) under pulsed laser excitation at 405 nm.

Excitation spectra of luminescence are shown in Fig. 7b. These spectra were recorded by controlling the wavelengths at 510 nm and 570 nm in the tails of emission bands in order to avoid their spectral overlap. The dominant contribution to the luminescence excitation spectra is made by the band with a maximum at 390 nm, which, in our opinion, is associated with optical transitions *def → def\** with the participation of local electronic states of surface defects (carbonyl and carboxyl groups) in CNDs. Less intense band located at 350 nm in excitation spectra is caused by σ → π\* optical transition. Note that the band associated with interband σ → σ\* optical transitions, which was well manifested in absorption spectra, is not detected in the luminescence excitation spectra. This fact indicates that when excited σ → σ\* transition, nonradiative energy transfer to the quenching centers is realized. The emission decay kinetics ($\lambda_{em}$ = 510 nm and $\lambda_{em}$ = 570 nm) under pulsed laser excitation at 405 nm are shown in Fig. 7c,d. The decay curves of both emission bands are well described by single exponent. The lifetimes of the excited states responsible for luminescence at 510 nm and 570 nm are determined to be 1.6 ± 0.1 ns and 0.9 ± 0.2 ns, respectively.

Comparison of the data obtained from excitation and emission spectra indicates the implementation of two "excitation – radiative relaxation" with the participation of band states (σ and π\*) and local energy levels of defects in the ground and excited states (*def* and *def\**):

- direct (intracenter) excitation of surface defect-related emission centers;
$$(def + h\nu_{exc} \rightarrow def^* \rightarrow def + h\nu_{em} + \hbar\omega)$$

- indirect excitation of surface defect-related emission centers involving band states;
$$(\sigma + h\nu_{exc} \rightarrow \pi^* \rightarrow def^* + \hbar\omega \rightarrow def + h\nu_{em} + \hbar\omega),$$

where $\hbar\omega$ is non-radiative energy losses. Figure 8 schematically shows the "excitation -relaxation" processes implemented in CDs fabricated from MLCNT.

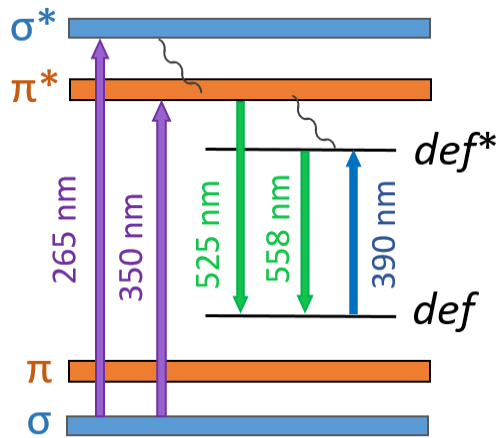

**Figure 8.** Scheme of "excitation - relaxation" processes implemented in CDs fabricated from MWCNTs. There are two channels for excitation of CDs luminescence with participation of band states (σ and π\*) and local energy levels of defects in the ground and excited states (*def* and *def\**).

In contrast to MWCNTs, laser treatment of SWCNTs leads to the formation of CDs with emission spectra typical for quantum dots (see Fig. 9a). Emission spectra of CQDs fabricated from SWCNTs are represented by narrow bands, the spectral position of which depends on the excitation wavelength. The nature of excitement-dependent luminescence in CDs is associated with quantum confinement effect (size effect) [65-67]. In this case, the emission is caused by the radiative recombination of an electron-hole pair in CDs of different sizes. Based on the spectral position of observed emission bands, we estimated the size of CDs using the equation proposed by Brus [68,69]:

$$E_{em} = E_g(bulk) + \frac{h^2}{8R^2}\left(\frac{1}{m_e^*} + \frac{1}{m_h^*}\right),  \quad (2)$$

where $E_{em}$ is the emission energy of CDs; $E_g(bulk)$ denotes the band gap of bulk material; $R$ is the radius of CDs; $m_e^*$ is the effective mass of electron; $m_h^*$ is the effective mass of hole; $h$ means the Planck's constant. In order to determine the radii of CDs we use experimental values $E_{em}$ obtained from emission spectra (Fig. 6a) and theoretical values $E_g(bulk) = 1.45$ eV [current work, see Section 3.2], $m_e^* = 0.57m_0$, $m_h^* = 0.8m_0$ (for diamond [70]). Results of calculations (see Fig. 9a) show the prevalence of CDs of 6.2 –13.3 Å in radius. A comparison of radius $R$ of nanodots with exciton Bohr radius $a_B$ allows to estimate the confinement regime: strong, intermediate or weak.

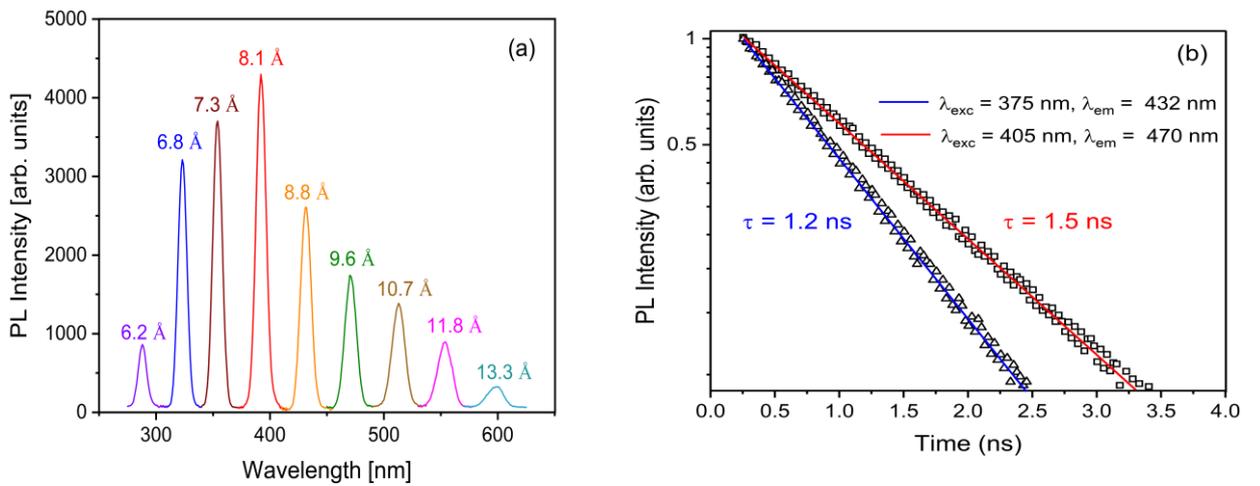

**Figure 9**. (a) Photoluminescence spectra ($\lambda_{exc} = 255 – 495$ nm) of CDs prepared from SWCNT. The numbers indicate the radii of CDs estimated from Eq. (1). (b) Decay kinetics of emission at 432 nm and 470 nm under pulsed laser excitation at 375 nm and 405 nm, respectively (approximation was performed with correlation coefficients $R^2 = 0.9854$ and $R^2 = 0.9679$, correspondingly).

Since the CDs under study, as we believe, have a diamene-like structure (see Sections 3.1 and 3.2), we use for comparison the exciton Bohr radius $a_B = 15.8$ Å for diamond [71]. Calculated values of CDs radii are less than the exciton Bohr radius which indicates the implementation of a strong confinement regime. We also note that, according to the emission decay kinetics (Fig. 9b), for smaller CDs, the lifetime of excited state takes a smaller value compared to larger CDs. This feature can be related with surface effects. Smaller CDs have an increased surface-to-volume ratio compared to larger CDs. So, a decrease in the size of QDs leads to an increase in the probability of non-radiative relaxations (due to the migration of excitation to the surface quenching centers) and, as a consequence, to a decrease in the lifetime of excited states [72].

Summarizing, we emphasize that CDs fabricated from MWCNTs and SWCNTs employing laser treatment show significantly different emission characteristics despite the similarity of their atomic structure. The reason for these differences seems to be related with the different size of the fabricated CDs. The evaluation demonstrated that the radius of the CNDs fabricated from SWCNTs varies in the range of 6.2 –13.3 Å. The size effect manifestation leads to excitation-dependent luminescence in a wide spectral range (from UV to red emission). On the contrary, laser fragmentation of MWCNTs leads to the formation of CNDs with sizes above dozens of nanometers. Such CNDs are characterized by excitation-independent luminescence in the green spectral range due to radiative transitions in surface oxygen-containing functional groups.

**4. Conclusions**

Based on the results of experimental study, modeling, and optical measurements, one can conclude that laser ablation of carbon nanotubes in water leads to the formation of 2D diamond-like carbon nanodots with oxidized surfaces and edges. In contrast to carbon nanodots fabricated by means of different methods, the studied samples demonstrate a high homogeneity of the atomic structure. The key difference between the synthesized carbon nanodots is that they all almost completely have *sp*$^3$ type of hybridization.

Based on XPS and Raman data, a theoretical model of two-dimensional diamond-like structure with the formula C$_2$OH was proposed. The simulation results also demonstrate the leading role of the edges in reducing band gap due to the contribution of oxygen-containing groups. On the contrary, the partial reduction of central parts of model C$_2$OH nanodots does not lead to the significant transformations of electronic structure. Optical absorption spectra of carbon nanodots fabricated from both sources are similar; therefore, based on the similarity of XPS group of spectra among each other and Raman group of spectra, one can assume the similarity of local atomic structure of carbon nanodots synthesized from SWCNTs and MWCNTs. The theoretical model developed by us assumes the interpretation of optical absorption spectra as a combination of σ → σ*, σ → π* transitions and some contribution from the edge states with adsorbed water.

The PL spectra of carbon nanodots fabricated from MWCNTs are in a good quantitative agreement with theoretical model of 2D diamond-like $C_2OH$ systems. At the same time, PL spectra of carbon nanodots fabricated from SWCNTs represent a set of distinct peaks typical for quantum dots. According to the simulations performed, the size of these carbon quantum dots was estimated as ~0.6–1.3 nm, and the largest quantity of these QCDs has a diameter of ~0.7–0.9 nm. Thus, there is every reason to assert that in our work a relatively simple and scalable method for the fabrication of carbon nanodots with a homogeneous structure and properties of quantum dots has been found.

## Acknowledgements

The work has been funded by the Russian Science Foundation (project № 21-12-00392). The equipment of the Ural Center for Shared Used "Modern Nanotechnologies" of Ural Federal University (Reg. 2968) was used for synthesis and characterization.